\documentclass[twocolumn,superscriptaddress,
showpacs,preprintnumbers,amsmath,amssymb]{revtex4}
\usepackage{graphicx}% Include figure files

\usepackage{epsfig}

\newcommand{\bra}[1]{{\left\langle #1 \right|}}
\newcommand{\ket}[1]{{\left| #1 \right\rangle}}

\newcommand{\Z}{\mbox{$\mathbb Z$}}
\newcommand{\cU}{\mathcal{U}}
\newcommand{\cS}{\mathcal{S}}
\newcommand{\cI}{\mathcal{I}}
\newcommand{\cW}{\mathcal{W}}
\newcommand{\cF}{\mathcal{F}}

\begin{document}
%%%%%%%%%%%%%%%%%%%%%%%%%%%%%%%%%%%%%%%%%%%%%%%%%%%%%%%%%%%%%%%%%%%%%%%%%%
%                                                                        %
%                                 Title                                  %
%                                                                        %
%%%%%%%%%%%%%%%%%%%%%%%%%%%%%%%%%%%%%%%%%%%%%%%%%%%%%%%%%%%%%%%%%%%%%%%%%%
\title{Quantum algorithms without initializing the auxiliary qubits}

\author{Dong Pyo Chi}\email{dpchi@math.snu.ac.kr}
\affiliation{
 School of Mathematical Sciences,
 Seoul National University, Seoul 151-742, Korea
}
\author{Jeong San Kim}\email{freddie1@snu.ac.kr}
\affiliation{
 School of Mathematical Sciences,
 Seoul National University, Seoul 151-742, Korea
}
\author{Soojoon Lee}\email{level@khu.ac.kr}
\affiliation{
 Department of Mathematics and Research Institute for Basic Sciences,
 Kyung Hee University, Seoul 130-701, Korea
}
\date{\today}

%%%%%%%%%%%%%%%%%%%%%%%%%%%%%%%%%%%%%%%%%%%%%%%%%%%%%%%%%%%%%%%%%%%%%%%%%%
%                                                                        %
%                              Abstract                                  %
%                                                                        %
%%%%%%%%%%%%%%%%%%%%%%%%%%%%%%%%%%%%%%%%%%%%%%%%%%%%%%%%%%%%%%%%%%%%%%%%%%
\begin{abstract}
In this Letter,
we construct the quantum algorithms for the Simon problem and the period-finding problem,
which do not require initializing
the auxiliary qubits involved in the process of functional evaluation
but are as efficient as the original algorithms.
In these quantum algorithms,
one can use any arbitrarily mixed state as the auxiliary qubits,
and furthermore can recover the state of the auxiliary qubits to the original one
after completing the computations.
Since the recovered state can be employed in any other computations,
we obtain that a single preparation of the auxiliary qubits in an arbitrarily mixed state
is sufficient to implement the iterative procedure
in the Simon algorithm or the period-finding algorithm.
\end{abstract}

%\keywords{}
\maketitle

%%%%%%%%%%%%%%%%%%%%%%%%%%%%%%%%%%%%%%%%%%%%%%%%%%%%%%%%%%%%%%%%%%%%%%
%%%                                                                %%%
%%%                         Introduction                           %%%
%%%                                                                %%%
%%%%%%%%%%%%%%%%%%%%%%%%%%%%%%%%%%%%%%%%%%%%%%%%%%%%%%%%%%%%%%%%%%%%%%
%%%
%%%     Teleportation and entanglement
%%%
Quantum computational algorithms can be executed in parallel on superpositions
of exponentially many input states,
and their outcomes can be properly measured by virtue of quantum interference.
These enable exponential speedups in the solutions of certain problems,
and allow one to distinguish
between the quantum computational complexity classes
and the classical ones \cite{DJ,BB92,BV,Simon,Shor,Grover,BBHT,BBBV,BH,Hallgren}.

It is assumed that most quantum algorithms require some initialization at start-up,
which is to prepare a certain pure state as an initial state.
However, in experimentally realizable proposals
for the implementation of quantum algorithms,
it may be technically difficult to prepare the initial pure state.
Especially, the nuclear magnetic resonance (NMR) system
is typically applied to physical systems in equilibrium at room temperature.
This means that
the initial state of the spins is nearly completely random,
that is,
it is difficult to prepare pure quantum states of nuclear spins in NMR systems.
Hence,
it would be interesting whether quantum algorithms can be efficiently performed
even though the initial states or some parts of them are not in a specific pure state.
If it would be possible, then the technical difficulty could be settled to a certain extent,
and furthermore if the parts of the initial state would remain intact even after the computation,
then the parts could be reused in any other computations.
We call such a quantum algorithm the {\it initialization-free quantum algorithm}
when any quantum state can be used as the auxiliary (target) qubits involved in the functional evaluation
$\ket{x} \otimes \ket{y} \mapsto \ket{x} \otimes \ket{y + f(x)}$ for a given function $f$,
and it can be recovered after the computation.

In the initialization-free quantum algorithms,
which are implemented without initializing
and deforming the state of the auxiliary qubits,
any qubits (which might contain some other useful information)
can be temporarily used as the auxiliary qubits, and
the initial state of the auxiliary qubits can be recovered
at the end of the computation.
Thus we can compose the auxiliary qubits of any qubits
regardless of whether they are entangled with others or
being used in another computational process.
Furthermore, in the case of iterative algorithms,
in which one needs to perform the algorithm several times to solve given problems,
the initialization-free quantum algorithms can be implemented with the same auxiliary qubits repeatedly,
while the original iterative algorithms require
the initial auxiliary qubits of a certain pure state at each repetition.

There have been a few research works related to the initialization-free quantum algorithms.
Biham {\it et al.}~\cite{Biham1, Biham2} have generalized Grover's algorithm~\cite{Grover}
by allowing for an arbitrary initial amplitude distribution,
and have shown that Grover's algorithm (or, a large class of Grover-type algorithms)
is robust against modest noise in the amplitude initialization procedure.
Parker and Plenio~\cite{PP}
found that one pure qubit and an initial supply of $\log_2 N$ qubits in an arbitrarily mixed state
are sufficient to implement Shor's quantum factoring algorithm \cite{Shor} efficiently,
where the idea of using one pure qubit and other mixed qubits as the initial state
was first introduced by Knill and Laflamme~\cite{Knill}.
Their result implies that the controlled unitary transformations in Shor's algorithm
can be implemented without any initialization of the auxiliary qubits,
while the auxiliary qubits cannot be left in the initial state.
Subsequently, Chi {\it et al.}~\cite{CKL2} have presented a quantum algorithm
to implement an oracle computing
$\left| x \right\rangle \mapsto e^{2\pi i f(x)/M} \left| x \right\rangle$
for $f:{\mathbb Z}_N \rightarrow {\mathbb Z}_M$
by making use of an oracle of the form
$\left| x \right\rangle \otimes \left| y \right\rangle \mapsto
\left| x \right\rangle \otimes \left| y + f(x) \right\rangle$
without setting the auxiliary qubits to a definite state
before the computation,
and have shown that generalized Deutsch-Jozsa algorithms can be implemented
without any initialization of the auxiliary qubits
when an oracle computing ${\cU}_f:\left| x \right\rangle \otimes \left| y \right\rangle \mapsto
\left| x \right\rangle \otimes \left| y + f(x) \right\rangle$
%or ${\cU}_f^{\oplus}:\left| x \right\rangle \otimes \left| y \right\rangle \mapsto
%\left| x \right\rangle \otimes \left| y \oplus f(x) \right\rangle$
is employed.

In this Letter,
we deal with two problems, %the Bernstein-Vazirani (BV) problem~\cite{BV},
the Simon problem~\cite{Simon} and the period-finding problem~\cite{Shor},
which can be solved efficiently by the quantum computer, and
present the initialization-free quantum algorithms for these problems.
Since most of known ``exponentially fast" applications of the quantum Fourier transform (QFT)
can be considered as a generalization of finding unknown period of a periodic function,
the existence of these quantum algorithms implies that any initialization of the auxiliary qubits
may be unnecessary in many quantum algorithms.

%%%%%%%%%%%%%%%%%%%%%%%%%%%%%%%%%%%%%%%%%%%%%%%%%%%%%%%%%%%%%%%%%%%%%%%%%
%%%                                                                   %%%
%%%                    The orginal Simon Algorithm                    %%%
%%%                                                                   %%%
%%%%%%%%%%%%%%%%%%%%%%%%%%%%%%%%%%%%%%%%%%%%%%%%%%%%%%%%%%%%%%%%%%%%%%%%%
We first recall the Simon problem~\cite{Simon}
that can be solved in polynomial time on a quantum computer
but that requires exponential time
on any classical bounded-error probabilistic Turing machine
if the data is supplied as a black box, and
we then investigate initialization-free techniques for the Simon problem.
We note that
there exists an exact quantum polynomial-time algorithm
for the Simon problem~\cite{BH}.
However, we here deal with the original Simon algorithm
which is polynomial-time in the expected sense.

For convenience, we use the following notations.
Let $G=\left({\mathbb Z}_2^n, \oplus_n\right)$ be a group under the binary operation $\oplus_n$,
which is the bitwise XOR operation.
For a subset $A$ of $G$, let $|A|$ denote the cardinality of $A$.
%Throughout this Letter, we assume that ${\mathbb Z}_2^n$ is identified with ${\mathbb Z}_{2^n}$ in the following sense:
%$(a_0, a_1, \cdots, a_{n-1})\leftrightarrow \sum_{j=0}^{n-1}a_j 2^j$.

We define a bilinear map $G\times G \rightarrow {\mathbb Z}_2$ by
\begin{equation}
x\cdot y=x_0y_0\oplus x_1y_1\oplus \cdots \oplus x_{n-1}y_{n-1}
\label{eq:bilinear}
\end{equation}
where %$x=\sum_{j=0}^{n-1}x_j 2^j$ and $y=\sum_{j=0}^{n-1}y_j 2^j$ are elements in $G$
$x=(x_0, x_1, \cdots, x_{n-1})$ and $y=(y_0, y_1, \cdots, y_{n-1})$ are elements in $G$
for $x_j, y_j \in {\mathbb Z}_2$,
and $\oplus$ is the XOR operation,
which is the addition modulo 2.
This bilinear map clearly satisfies the property that
$(x\oplus_n y)\cdot z=x\cdot z \oplus y\cdot z$
for $x, y$ and $z$ in $G$.

For a subgroup $H$ of $G$, let
\begin{equation}
H^{\perp}=\{x\in G: x\cdot y=0 ~\textrm{  for all  } y\in H\}
\label{eq:perp}
\end{equation}
denote the orthogonal subgroup of $H$.
We remark that
the quotient group $G/H$ is well-defined
since $G$ is an abelian group.

Let $f:G \rightarrow G$ be an arbitrary two-to-one map such that
$f(x)=f(y)$ if and only if $x\oplus_n y\in H$
where $H=\{0,h\}$ is a subgroup of $G$ for some nonzero $h\in G$.
Then the Simon problem is to find the subgroup $H$, that is, to determine the value of $h$.
The original Simon algorithm is %similar to the BV algorithm
as follows:
(i)~Prepare $\left|{0^n}\right\rangle \otimes \left|{0^n}\right\rangle$.
(ii)~Apply ${\cW}_n \otimes {\cI}$,
where ${\cW}_n$ is the $n$-qubit Walsh-Hadamard transform
defined as $\ket{x}\mapsto (1/\sqrt{|G|})\sum_{y\in G}(-1)^{x\cdot y}\ket{y}$.
(iii)~Apply ${\cU}_f$.
(iv)~Apply ${\cW}_n \otimes {\cI}$.
Then the resulting state
is
\begin{equation}
\left|{\Phi}\right\rangle=\frac{2}{|G|} \sum_{y\in{H^{\perp}}}\sum_{\bar{x}\in G/H}
 (-1)^{x\cdot y}\left|{y}\right\rangle
 \otimes  \left|{f(x)}\right\rangle.
\label{eq:BV_result}
\end{equation}
We measure the first $n$-qubit state. Then
for each $y\in H^{\perp}$,
the probability with which we obtain $y$ as the outcome is
\begin{equation}
\left\langle{\Phi}\right|\left(\left|{y}\right\rangle \left\langle{y}\right|
\otimes {\cI}\right)\left|{\Phi}\right\rangle
= \frac{4}{|G|^2}\sum_{\bar{x}\in G/H} 1
=\frac{2}{|G|}.
\label{eq:BV_prob}
\end{equation}
Thus, after expected $O(n)$ repetitions of this procedure,
at least $n$ linearly independent values of $y$ can be collected
so that the nontrivial $h^*$ is uniquely determined
by solving the linear system of equations $h^*\cdot y=0$.
Then we have $h^*=h$ as required.

%%%%%%%%%%%%%%%%%%%%%%%%%%%%%%%%%%%%%%%%%%%%%%%%%%%%%%%%%%%%%%%%%%%%%%%%%
%%%                                                                   %%%
%%%                Initialization-free Simon Algorithm                %%%
%%%                                                                   %%%
%%%%%%%%%%%%%%%%%%%%%%%%%%%%%%%%%%%%%%%%%%%%%%%%%%%%%%%%%%%%%%%%%%%%%%%%%
Now we present the initialization-free quantum algorithm for the Simon problem.
We first consider the following quantum algorithm:
(i)~Prepare an $n$-qubit state in the state $\ket{0^n}$ as the control qubits
and an $n$-qubit state in an arbitrary pure state $\ket{\Psi}=\sum_{k}\alpha_k \ket{k}$
as the auxiliary qubits.
(ii)~Apply ${\cW}_n \otimes {\cI}$.
(iii)~Apply ${\cU}_f^{\oplus}$,
where ${\cU}_f^{\oplus}:\left| x \right\rangle \otimes \left| y \right\rangle \mapsto
\left| x \right\rangle \otimes \left| y \oplus_n f(x) \right\rangle$.
(iv)~Choose a random $n$-bit string $w=(w_0 , w_1 ,\cdots , w_{n-1} )$
and apply $\cS_w = \sigma_z^{w_0}\otimes\sigma_z^{w_1} \otimes \cdots \otimes \sigma_z^{w_{n-1}}$
on the $n$-qubit auxiliary qubits, that is, apply $\cI \otimes \cS_w$.
(v)~Apply ${\cU}_f^{\oplus}$.
(vi)~Apply $\cI \otimes \cS_w$.
(vii)~Apply ${\cW}_n \otimes {\cI}$.
Then %the state evolves as follows:
the resulting state becomes
\begin{eqnarray}
 \frac{2}{|G|}
\sum_{y\in H^{\perp}}\left(\sum_{\bar{x}\in G/H }
(-1)^{w\cdot f(x)}(-1)^{x\cdot y }\right)\ket{y}\otimes\ket{\Psi}.
  \label{Alg:IFSimon}
\end{eqnarray}
We now measure the first $n$-qubit state. Then for each $y \in H^{\perp}$,
the probability with which we obtain $y$ as the measurement outcome is
\begin{equation}
{P}_{w}(y) =\frac{4}{|G|^2} \left|\sum_{\bar{x}\in G/H
}(-1)^{w\cdot f(x)}(-1)^{x\cdot y }\right|^2.
   \end{equation}
Hence, the expected probability of obtaining $y$ for
%Here, we take the average on
randomly chosen $w$ is
%Then
\begin{widetext}
\begin{eqnarray}
\frac{1}{|G|}\sum_{w\in G}{P}_{w}(y)
&=&
\frac{4}{|G|^3}\sum_{w\in G}
\left|\sum_{\bar{x}\in G/H}(-1)^{w\cdot f(x)}(-1)^{x\cdot y }\right|^2
\nonumber \\
&=&
\frac{4}{|G|^3}\sum_{\bar{x}, \bar{x}'\in G/H}\left(\sum_{w\in G }(-1)^{w\cdot (f(x)\oplus_n f(x'))}\right)
(-1)^{(x \oplus_n x') \cdot y}.
\label{eq:average_Simon}
\end{eqnarray}
\end{widetext}
Since $f$ is one-to-one on $G/H$, that is, $f(x)\neq f(x')$ if and only if $\bar{x}\neq\bar{x}'$,
the inner summation in (\ref{eq:average_Simon})
always vanishes for $f(x)\oplus_n f(x') \neq 0$,
and the summation is $|G|$ for $f(x)\oplus_n f(x') = 0$.
Thus, for each $y \in H^{\perp}$,
the expected probability (\ref{eq:average_Simon}) becomes $2/|G|$.

%%%%%%%%%%%%%%%%%%%%%%%%%%%%%%%%%%%%%%%%%%%%%%%%%%%%%%%%%%%%%%%%%%%%%%%%%
%%%                                                                   %%%
%%%             Mixed state and superoperator(Simon)                  %%%
%%%                                                                   %%%
%%%%%%%%%%%%%%%%%%%%%%%%%%%%%%%%%%%%%%%%%%%%%%%%%%%%%%%%%%%%%%%%%%%%%%%%%
For any auxiliary qubits of the state $\rho_B = \sum_{k} p_k \ket{\Psi_k}\bra{\Psi_k}$,
we let $\rho = \ket{0^n}\bra{0^n} \otimes \rho_B$.
Then the superoperator $\Lambda$, which maps $\rho$ to the quantum state
\begin{equation}
\frac{1}{|G|}\sum_{w \in G} \Lambda_w \rho \Lambda_w^\dagger
\label{super}
 \end{equation}
where $ \Lambda_w = (\cW_n \otimes
\cI)(\cI\otimes\cS_w)\cU_f^{\oplus}(\cI\otimes\cS_w)
 \cU_f^{\oplus}(\cW_n \otimes \cI)$,
performs the initialization-free Simon algorithm,
since
if the first $n$-qubit state in $\Lambda(\rho)$ is measured,
then it follows from (\ref{eq:average_Simon})
that the probability to obtain $y$ as the measurement outcome
is
\begin{eqnarray}
 \mathrm{tr}\left[ (\ket{y} \bra{y} \otimes \cI)\Lambda (\rho)\right]
 &=& \frac{1}{|G|}\sum_{w\in G}{P}_{w}(y)
 = \frac{2}{|G|},
\end{eqnarray}
which is the same probability as that of the original Simon algorithm.
Furthermore, when $y$ is obtained as the measurement outcome,
the resulting state after the measurement becomes
$\ket{y}\bra{y}\otimes\rho_B$.
Therefore, this initialization-free quantum algorithm can efficiently solve the Simon problem.

%%%%%%%%%%%%%%%%%%%%%%%%%%%%%%%%%%%%%%%%%%%%%%%%%%%%%%%%%%%%%%%%%%%%%%%%%
%%%                                                                   %%%
%%%                   The Period Finding Algorithm                    %%%
%%%                                                                   %%%
%%%%%%%%%%%%%%%%%%%%%%%%%%%%%%%%%%%%%%%%%%%%%%%%%%%%%%%%%%%%%%%%%%%%%%%%%
Similarly, we can present the initialization-free quantum algorithm for the period-finding problem.
We first review the original quantum algorithm for the period-finding problem,
and then present the initialization-free period-finding algorithm,
which can be considered as a generalization of the initialization-free Simon algorithm.

Let $f: \Z_{2^n} \rightarrow \Z_{2^m}$ be a periodic function
with an unknown period $T$, that is,
$f(x) = f(x+kT)$ for $0\le k\le \lfloor 2^n/T\rfloor$ (or, $0\le k\le\lfloor 2^n/T\rfloor+1$).
Then the period-finding problem is to find $T$.
 Classically, this problem is known to be hard in the sense that
 no classical algorithms which can find $T$ in polynomial time have been found.
However, there exists a polynomial-time quantum algorithm for the period finding~\cite{Shor},
which is as follows: Let $N=2^n$.
(i) Prepare $\left|{0^n}\right\rangle \otimes \left|{0^m}\right\rangle$.
(ii) Apply ${\cF} \otimes {\cI}$,
where $\cF$ is the $N$-dimensional QFT
defined as $\ket{x}\mapsto (1/\sqrt{N})\sum_{y=0}^{N-1}e^{2\pi i xy/N}\ket{y}$.
(iii) Apply ${\cU}_f$.
(iv) Apply ${\cF} \otimes {\cI}$.
Then the resulting state becomes
\begin{eqnarray}
 \frac{1}{N} \sum_{y=0}^{N-1}\sum_{x=0}^{T-1} \sum_{j=0}^{A_x-1} e^{2\pi i y(x+jT)/N} \ket{y}
 \otimes   \ket{ f(x) }
   \label{Alg:Period}
\end{eqnarray}
where $A_{x}= \lfloor N/T \rfloor$ or $ \lfloor N/T \rfloor + 1$.
Now we measure the first $n$-qubit state,
and then the probability of obtaining $y$ as a measurement outcome is
\begin{equation}
P(y) =
\frac{1}{N^2}\sum_{x=0}^{T-1}\left| \sum_{j=0}^{A_x} e^{2\pi i yjT/N}\right|^2.
\label{eq:py}
\end{equation}
We note that there are precisely $T$ values of $y$ in
$\{0,1,\cdots ,N-1\}$ satisfying
\begin{eqnarray}
-\frac{T}{2} \leq yT~(\mathrm{mod}~N)\leq \frac{T}{2},
\label{eq:yr}
\end{eqnarray}
and for each $y$ satisfying (\ref{eq:yr}), the probability of obtaining
such $y$ can be bounded asymptotically,
\begin{equation}
\textrm{P}(y) \geq \frac{4}{\pi^2} \frac{1}{T}.
\label{eq:final_Py}
\end{equation}
Thus, with probability at least $4/\pi^2$, the measured value of $y$ satisfies the inequalities~(\ref{eq:yr}),
that is, $y$ satisfies the following inequalities:
\begin{equation}
\frac{k}{T}- \frac{1}{2N}\leq\frac{y}{N} \leq \frac{k}{T}+ \frac{1}{2N},
\label{eq:y_cond1}
\end{equation}
or equivalently
\begin{equation}
\left| \frac{y}{N}-\frac{k}{T}\right| \leq  \frac{1}{2N}
\label{eq:y_cond2}
\end{equation}
with $k$ randomly chosen in $\left\{0,1, \cdots ,T-1  \right\}$ depending on the measurement outcome.
Therefore, for sufficiently small $T$ with respect to $N$,
the value $k/T$ can be efficiently extracted from the measured $y/N$ by the continued fraction method.
Since $k$ and $T$ may be relatively prime with high probability,
we can get the period $T$ in polynomial time with respect to $\log{N}$.

The initialization-free quantum algorithm for the period-finding problem
can be presented by the procedure similar to the initialization-fee Simon algorithm.
Instead of $\cS_w = \sigma_z^{w_0}\otimes\sigma_z^{w_1} \otimes \cdots \otimes \sigma_z^{w_{n-1}}$
for a randomly chosen $n$-bit string $w$ in (\ref{Alg:IFSimon}),
we employ an $m$-qubit unitary operation $\cU_w$
for a randomly chosen $m$-bit string $w$
defined as
$\ket{y}\mapsto e^{2\pi i w y/M}\ket{-y}$,
where $M=2^m$.
We proceed with the following quantum algorithm:
(i)~Prepare an $n$-qubit state in the state $\ket{0^n}$
and an $m$-qubit state in an arbitrary pure state $\ket{\Psi}=\sum_{k}\alpha_k \ket{k}$.
(ii)~Apply ${\cF} \otimes {\cI}$.
(iii)~Apply ${\cU}_f$.
(iv)~Choose a random $m$-bit string $w$
and apply $\cU_w$ on the $m$-qubit state of the auxiliary qubits,
that is, apply $\cI \otimes \cU_w$.
(v)~Apply ${\cU}_f$.
(vi)~Apply $\cI \otimes \cU_w$.
(vii)~Apply ${\cF} \otimes {\cI}$.
Then the resulting state becomes
\begin{eqnarray}
 \frac{1}{N} \sum_{y=0}^{N-1} \left(\sum_{x=0}^{T-1} \sum_{j=0}^{A_x -1} e^{2\pi i y %\cdot
(x+jT)/N} e^{2\pi i w %\cdot
f(x)/M}\right)\ket{y}
 \otimes \ket{\Psi}.
   \nonumber \\
  \label{Alg:period}
\end{eqnarray}
Hence, the probability with which we can get $\ket{y}$
as a measurement outcome of the first $n$-qubit state is
\begin{eqnarray}
{P}_{w}(y) &=&\frac{1}{N^2}
\left|
\sum_{x=0}^{T-1}
\sum_{j=0}^{A_x -1}
e^{2\pi i y  (x+jT)/N}
e^{2\pi i w  f(x)/M}
\right|^2.
\nonumber \\
\label{eq:period_prob1}
\end{eqnarray}
By straightforward calculations, we can get
the expected probability of obtaining $y$ for randomly chosen $w$,
\begin{eqnarray}
\frac{1}{M}\sum_{w=0}^{M-1}{P}_{w}(y)
&=&
\frac{1}{N^2}\sum_{x=0}^{T-1}\left| \sum_{j=0}^{A_x} e^{2\pi i yjT/N}\right|^2.
\label{eq:period_prob3}
\end{eqnarray}

%%%%%%%%%%%%%%%%%%%%%%%%%%%%%%%%%%%%%%%%%%%%%%%%%%%%%%%%%%%%%%%%%%%%%%%%%
%%%                                                                   %%%
%%%             Mixed state and superoperator(period finding)         %%%
%%%                                                                   %%%
%%%%%%%%%%%%%%%%%%%%%%%%%%%%%%%%%%%%%%%%%%%%%%%%%%%%%%%%%%%%%%%%%%%%%%%%%

As in the initialization-free Simon algorithm,
for any $m$-qubit state $\rho_B= \sum_{k} p_k \ket{\Psi_k}\bra{\Psi_k}$,
we let $\rho = \ket{0^n}\bra{0^n} \otimes \rho_B$,
and let the superoperator $\Lambda$ be defined as
\begin{equation}
\rho\mapsto \frac{1}{M}\sum_{w =0}^{M-1} \Lambda_w \rho \Lambda_w^\dagger
\label{eq:PF_super}
 \end{equation}
where $ \Lambda_w = (\cF \otimes\cI)(\cI\otimes\cU_w)\cU_f(\cI\otimes\cU_w)
 \cU_f(\cF \otimes \cI)$.
Then the superoperator $\Lambda$ can
perform the period-finding algorithm efficiently without any
initialization on the auxiliary qubits, since the probability of
obtaining $\ket{y}$ satisfying (\ref{eq:yr}) is
\begin{eqnarray}
\mathrm{tr} \left[(\ket{y} \bra{y} \otimes \cI)\Lambda (\rho) \right]
= \frac{1}{N^2}\sum_{x=0}^{T-1}\left| \sum_{j=0}^{A_x} e^{2\pi i yjT/N}\right|^2,
\label{eq:last}
 \end{eqnarray}
which is the same probability as that of the original period-finding algorithm in (\ref{eq:py}).
Furthermore, as in the initialization-free Simon algorithm,
the resulting state after the measurement becomes
$\ket{y}\bra{y}\otimes\rho_B$ when $y$ is the measurement outcome.
Therefore, there exists an initialization-free quantum algorithm
which can efficiently solve the period-finding problem.

%%%%%%%%%%%%%%%%%%%%%%%%%%%%%%%%%%%%%%%%%%%%%%%%%%%%%%%%%%%%%%%%%%%%%%
%%%                                                                %%%
%%%                         Conclusion                             %%%
%%%                                                                %%%
%%%%%%%%%%%%%%%%%%%%%%%%%%%%%%%%%%%%%%%%%%%%%%%%%%%%%%%%%%%%%%%%%%%%%%
In conclusion,
we have investigated the initialization-free quantum algorithms,
which do not require any initialization of the auxiliary qubits
involved in the process of functional evaluation,
and which recover the initial state of the auxiliary qubits
after completing the computations.
We have considered quantum algorithms for the Simon problem and the period-finding problem,
and have presented the initialization-free quantum algorithms for the problems,
which are as efficient as the original ones.

The iterative algorithms such as the Simon algorithm and the period-finding algorithm
demand the storage of auxiliary qubits and
the extra operations to initialize the state of the auxiliary qubits,
whenever the procedure repeats.
However if one utilizes our initialization-free technique then
the size of the storage can be reduced and the extra operations can be omitted,
since the same auxiliary qubits can repeatedly be used in our algorithms.
Furthermore, since most known applications of the QFT
can be considered as a generalization of finding unknown period of a periodic function
(for example, Shor's factoring algorithm~\cite{Shor}
and Hallgren's more recent algorithm for solving Pell's equation~\cite{Hallgren}),
the initialization-free technique could be applied to a lot of implementations of quantum algorithms.

%%%%%%%%%%%%%%%%%%%%%%%%%%%%%%%%%%%%%%%%%%%%%%%%%%%%%%%%%%%%%%%%%%%%%%
%%%                                                                %%%
%%%                       Acknowledgements                         %%%
%%%                                                                %%%
%%%%%%%%%%%%%%%%%%%%%%%%%%%%%%%%%%%%%%%%%%%%%%%%%%%%%%%%%%%%%%%%%%%%%%
The authors would like to thank D.~A.~Lidar and M.~Grassl for helpful comments.
D.P.C. was supported by
a Korea Research Foundation Grant (KRF-2004-059-C00060)
and by Asian Office of Aerospace Research and Development (AOARD-04-4003).

\end{document}